# Cyclooxygenase Inhibition Limits Blood-Brain Barrier Disruption following Intracerebral Injection of Tumor Necrosis Factor-α in the Rat

Eduardo Candelario-Jalil, Saeid Taheri, Yi Yang, Rohit Sood, Mark Grossetete, Eduardo Y. Estrada, Bernd L. Fiebich, Gary A. Rosenberg

*Department of Neurology, University of New Mexico Health Sciences Center, Albuquerque, NM 87131, USA* **(ECJ, YY, MG, EYE, GAR)**

*MRI Core, Biomedical Research and Integrative NeuroImaging (BRaIN) Center, University of New Mexico, Albuquerque, NM 87131, USA* **(ST, RS)**

*Neurochemistry Research Group, Department of Psychiatry, University of Freiburg Medical School, Freiburg D-79104, Germany* **(BLF)**

*Running Title:* Cyclooxygenase inhibition reduces TNF-α-induced BBB opening

*Corresponding author:*   **Eduardo Candelario-Jalil, PhD**
Department of Neurology, University of New Mexico Health Sciences Center, MSC10 5620, Albuquerque, NM 87131-0001
USA
Tel: (505)-925-4042. Fax: (505)-272-6692.
Email: ECandelario-Jalil@salud.unm.edu

*Abbreviations:* ADC, Apparent Diffusion Coefficient; BBB, blood-brain barrier; COX, cyclooxygenase; DMSO, dimethyl sulfoxide; DWI, Diffusion Weighted Imaging; FRET, fluorescence resonance energy transfer; GFAP, glial fibrillary acidic protein; GSH, reduced glutathione; LPS, lipopolysaccharide; MCB, monochlorobimane; MMP, matrix metalloproteinase; MRI, magnetic resonance imaging; NeuN, neuron-specific nuclear protein; ROI, region of interest; TNF-α, tumor necrosis factor-α; VAS, Valeroyl Salicylate

**ABSTRACT**

Increased permeability of the blood-brain barrier (BBB) is important in neurological disorders. Neuroinflammation is associated with increased BBB breakdown and brain injury. Tumor necrosis factor-$\alpha$ (TNF-$\alpha$) is involved in BBB injury and edema formation through a mechanism involving matrix metalloproteinase (MMP) upregulation. There is emerging evidence indicating that cyclooxygenase (COX) inhibition limits BBB disruption following ischemic stroke and bacterial meningitis, but the mechanisms involved are not known. We used intracerebral injection of TNF-$\alpha$ to study the effect of COX inhibition on TNF-$\alpha$-induced BBB breakdown, MMP expression/activity and oxidative stress. BBB disruption was evaluated by the uptake of $^{14}$C-sucrose into the brain and by magnetic resonance imaging (MRI) utilizing Gd-DTPA as a paramagnetic contrast agent. Using selective inhibitors of each COX isoform, we found that COX-1 activity is more important than COX-2 in BBB opening. TNF-$\alpha$ induced a significant upregulation of gelatinase B (MMP-9), stromelysin-1 (MMP-3) and COX-2. In addition, TNF-$\alpha$ significantly depleted glutathione as compared to saline. Indomethacin (10 mg/kg; i.p.), an inhibitor of COX-1 and COX-2, reduced BBB damage at 24 h. Indomethacin significantly attenuated MMP-9 and MMP-3 expression and activation, and prevented the loss of endogenous radical scavenging capacity following intracerebral injection of TNF-$\alpha$. Our results show for the first time that BBB disruption during neuroinflammation can be significantly reduced by administration of COX inhibitors. Modulation of COX in brain injury by COX inhibitors or agents modulating prostaglandin E$_2$ formation/signaling may be useful in clinical settings associated with BBB disruption.

**INTRODUCTION**

Increased permeability of the BBB is a pathological hallmark in several neurological disorders (Hawkins and Davis, 2005). Neuroinflammatory events are associated with increased BBB breakdown and brain injury. Tumor necrosis factor-$\alpha$ (TNF-$\alpha$) has been demonstrated to participate in the BBB opening, which follows cerebral ischemic injury (Yang et al., 1999), sepsis (Tsao et al., 2001), bacterial meningitis (Ramilo et al., 1990), and human immunodeficiency virus type-1 (HIV-1) infection (Fiala et al., 1997). TNF-$\alpha$-mediated BBB breakdown is linked to the upregulation of matrix metalloproteinase (MMP) expression and activity (Rosenberg et al., 1995b).

Metabolism of arachidonic acid through cyclooxygenase (COX) plays a key role in neuroinflammatory events (Hewett et al., 2006). COX inhibition effectively blocked BBB permeability increases induced by TNF-$\alpha$ in an *in vitro* model of BBB using bovine brain microvessel endothelial cells (BBMECs) (Mark et al., 2001c). They showed that the increased permeability and cytoskeletal structural changes observed in the BBMECs following exposure to TNF-α involved the induction of COX-2 with increased release of prostaglandins (Mark et al., 2001b). Prostaglandin $E_2$ ($PGE_2$), the major prostanoid produced by COX activity in the brain, produced marked BBB breakdown when administered intracerebrally in the rat (Schmidley et al., 1992). Selective COX-2 inhibition with NS-398 reduced the effects of TNF-α on cerebromicrovascular permeability in a rat cranial window model (Trickler et al., 2005). Furthermore, COX inhibition with indomethacin significantly reduced BBB disruption induced by TNF-α *in vitro* (de Vries et al., 1996).

COX inhibition reduces the expression of MMP-2 and MMP-9 in human monocytes (Cipollone et al., 2005), macrophages (Pavlovic et al., 2006), and several tumor cell lines (Pan et al., 2001; Kinugasa et al., 2004). Additionally, COX inhibition has been proven to reduce oxidative stress during neuroinflammation (Candelario-Jalil et al., 2003d; Akundi et al., 2005b), which might also contribute to the reduction of MMP expression and activity seen in peripheral cells. We recently showed that the COX-2 inhibitor, nimesulide, reduces BBB damage and leukocyte infiltration following focal cerebral ischemia (Candelario-Jalil et al., 2007b). However, the precise molecular mechanism(s) through which COX inhibition limits BBB opening, and its possible relation to MMP expression/activity, had not been previously investigated.

In the study presented here, we utilized a well-characterized animal model of TNF-α-induced BBB disruption to investigate the effects of COX inhibition on MMP expression and activity. By using selective inhibitors of COX-1 and COX-2, we studied the relative contribution of each isoform to TNF-α-mediated BBB opening. We showed that COX-1 selective inhibition or inhibition of both isozymes with indomethacin significantly reduced BBB damage in this model of neuroinflammation, as assessed by the $^{14}$C-sucrose uptake method and corroborated by MRI permeability measurements (Sood et al., 2007a). Inhibition of COX with indomethacin reduced the significant increase in MMP-3 and MMP-9 expression/activity and prevented the loss of endogenous antioxidant capacity induced by TNF-α. Our present results indicate that COX is important in BBB permeability changes induced by TNF-α. Upregulation of MMPs and increase in oxidative stress link COX activity to BBB disruption.

MATERIALS AND METHODS

**Stereotaxic Surgery and Intracerebral Administration of TNF-α in the Rat**

All animal studies were approved by the University of New Mexico Health Sciences Center's Animal Care and Use Committee and conformed to NIH guidelines for use of animals in research. Male Wistar rats (Harlan, Indianapolis, IN) weighing 280-320 g were used in this study. Animals were anesthetized with 1.5% halothane in 70% nitrous oxide and 30% oxygen. Rats were then positioned in a stereotaxic headholder (Kopf Instruments, Tujunga, CA) and a burr hole was made in the skull 3 mm lateral from midline, at bregma. Intrastriatal injections of rhTNF-α (5000 U of recombinant human TNF-α; Upstate, Lake Placid, NY) in 10 μl of sterile physiological saline (Hospira Inc., Lake Forest, IL) were administered 5 mm below the *dura* over a 10 min duration with a microsyringe (Hamilton, Reno, NV). The needle was left in place another 3 minutes. Evan's blue dye was added to the infusate for localization of the injection site. The burr hole was sealed with bone wax and the skin incision was closed with 4-0 silk sutures.

**Evaluation of BBB Permeability with $^{14}$C-Sucrose**

Opening of the BBB induced by TNF-α was evaluated using the sucrose space method, as described in detail previously (Rosenberg et al., 1998a). Briefly, animals were anesthetized and infused through the femoral vein with 10 μCi of $^{14}$C-sucrose (molecular mass 342 Da; PerkinElmer, Waltham, MA). Ten minutes later, a sample of blood was drawn from the femoral vein, and animals were sacrificed with an intracardiac injection of a saturated KCl solution. Brains were rapidly removed and frozen in 2-methylbutane cooled to -80°C in liquid nitrogen. For evaluating the BBB breakdown, striata and overlying cortex were dissected out for regional evaluation of BBB opening. Tissues were dissolved in Solvable (DuPont, Boston, MA) and $^{14}$C-sucrose was measured in both brain samples and plasma in a liquid scintillation counter (Beckman LS6500 Scintillation Counter; GMI Inc., Ramsey, MN) using CytoScint as the liquid scintillation cocktail (ICN Biomedicals, Costa Mesa, CA). Ratios were used to calculate brain sucrose uptake as an estimate of capillary permeability (Rosenberg et al., 1998b).

**Gelatin-Substrate Zymography**

For zymography, tissues were homogenized in a lysis buffer containing 50 mM Tris-HCl pH 7.6, 150 mM NaCl, 5 mM $CaCl_2$, 0.05% Brij-35, 1% Triton X-100, and 0.02% $NaN_3$. Protein concentration in the homogenate was determined using the Micro BCA Protein Assay kit (Pierce, Rockford, IL). MMP-2 and MMP-9 present in the homogenates were concentrated with Gelatin-Sepharose 4B beads (GE Healthcare Biosciences, Piscataway, NJ), and analyzed by gelatin zymography as previously described (Zhang and Gottschall, 1997). Briefly, 800 μg of total protein in 500 μl was incubated with 80 μl of Gelatin-Sepharose 4B beads for 1 h at 4°C with gentle agitation. The beads were collected and the gelatinases were eluted by incubating with 80 μl of elution buffer (10% DMSO in phosphate buffered saline) for 30 min at 4°C with gentle shaking. Equal amount of samples (20 μl) were mixed in a loading buffer containing 80 mM Tris-HCl pH 6.8, 4% sodium dodecyl sulfate (SDS), 10% glycerol, and 0.01% bromophenol blue. Samples were electrophoretically separated on 10% SDS-polyacrylamide gels co-polymerized with 1 mg/ml gelatin (Sigma-Aldrich, St. Louis, MO) under non-reducing conditions. Gels were washed in 2.5% Triton X-100 and then incubated for 40 h at 37°C in a developing buffer containing 50 mM Tris pH 7.6, 5 mM $CaCl_2$, 0.2 mM NaCl, and 0.02% Brij-35. After incubation, gels were stained with 0.125% Coomassie Brilliant Blue R-250 (Sigma-Aldrich, Saint Louis, MO) for 30 min in 10% acetic acid and 50% methanol. Gels were destained with a solution containing 10% acetic acid and 10% methanol until clear bands of gelatinolysis appeared on a dark blue background. To confirm that detected activities were zinc-dependent gelatinases, some zymogram gels were incubated with developing solution in the presence of 10 mM EDTA. Disappearance of lytic bands in these gels confirmed the metal dependence of gelatinolytic activity characteristic of MMPs. The gels were dried and densitometrically scanned (Alpha Imager$^{TM}$ 2200; Alpha Innotech, San Leandro, CA). Molecular weights were determined by both protein standards (Bio-Rad Laboratories, Hercules, CA), and conditioned media from HT1080 human fibrosarcoma cells, a well-known source of MMP-2 and -9. In addition, a mixture of active MMP-2 and -9 standards (Chemicon International, Temecula, CA) were used as gelatinase controls. Data were expressed as relative lysis units.

**Immunoblotting**

We added to part of the homogenate prepared for zymography complete protease inhibitor cocktail (Roche Diagnostics GmbH, Mannheim, Germany). Before electrophoresis, samples (60 μg of total protein) were mixed 1:1 with Laemmli sample buffer (65 mM Tris-HCl pH 6.8, 10% glycerol, 2% SDS, 0.002% bromophenol blue) containing 5% of 2-mercaptoethanol. Samples were incubated at 100°C for 10 min and subjected to SDS-PAGE on a 7.5% gel (for MMP-9, MMP-2 and COX-2) or 10% gel (for MMP-3 and COX-1) under reducing conditions. Proteins were then transferred onto a polyvinylidene fluoride (PVDF) membrane (Millipore, Bedford, MA) by wet blotting. The membrane was blocked for 1 h at room temperature with 5% skim milk in Tris-buffered saline (TBS) containing 0.1% Tween 20 (TBST) before incubation with the primary antibody. Primary polyclonal antibodies were rabbit anti-MMP-9 (AB19016; Chemicon International, Temecula, CA; 1:1500), rabbit anti-MMP-2 (sc-10736; Santa Cruz Biotechnology, Santa Cruz, CA; 1:200), rabbit anti-COX-2 (160126; Cayman Chemical, Ann Arbor, MI; 1:200), rabbit anti-COX-1 (160109; Cayman Chemical, Ann Arbor, MI; 1:2000) and goat anti-MMP-3 (ab18898; Abcam, Cambridge, MA; 1 μg/ml). Antibody dilutions were made in 5% milk dissolved in TBST. Membranes were incubated overnight at 4°C with the corresponding primary antibody. After extensive washing (three times for 15 min each in TBST), proteins were detected with horseradish peroxidase (HRP)-linked goat anti-rabbit IgG (#7074; Cell Signaling, Danvers, MA; 1:2000) or donkey anti-goat IgG (705-035-147; Jackson ImmunoResearch Laboratories, West Grove, PA; 1:10,000) using SuperSignal® West Pico Chemiluminescent substrate (Pierce, Rockford, IL). Each blot was stripped and reprobed with a rabbit anti-actin IgG (Sigma-Aldrich, Saint Louis, MO; 1:7500) to confirm equal protein loading and transfer. Quantification of the Western blots was performed using ImageJ software (National Institutes of Health; Bethesda, MD).

**Tissue Processing for Immunohistochemistry**

Twenty four hours after rhTNF-$\alpha$ or saline injection, rats were anesthetized with pentobarbital (50 mg/kg; i.p.), and transcardially perfused with cold phosphate buffered saline (PBS) followed by 2% PLP (2% paraformaldehyde, 0.1 M sodium periodate, 75 mM lysine in 100 mM sodium phosphate buffer at pH 7.4). Brains were removed and immersed in 2% PLP overnight. The brains were cryoprotected in 30% sucrose in 2% PLP, placed in a Peel-A-Way® histology mold (Ted Pella Inc., Redding, CA) containing Tissue-Tek® embedding medium (Sakura Finetek, Torrance, CA), and frozen with 2-methylbutane cooled in liquid nitrogen. Brains were stored at -80°C until sectioned. Tissue was sectioned on a cryostat (Richard-Allan Scientific, Kalamazoo, MI), and SuperFrost®/Plus slides (Fisher Scientific, Pittsburgh, PA) with 10 $\mu$m sections (5 sections per slide) were stored at -20°C until immunostaining was performed.

**Double Immunofluorescence Labeling and Confocal Microscopy**

Immunolabeling was performed for MMP-9, COX-2 and MMP-3. For investigating the cell types expressing each of these proteins following rhTNF-$\alpha$ injection, double immunolabeling was performed with specific cell markers, including glial fibrillary acidic protein (GFAP; marker of reactive astrocytes), neuron-specific nuclear protein (NeuN; marker of neurons) and OX-42 (also known as CD11b; marker of microglia/macrophages). Primary antibodies and dilutions used in the immunohistochemical analysis were: rabbit anti-MMP-9 (AB19016; Chemicon International, Temecula, CA; 1:300), rabbit anti-COX-2 (160126; Cayman Chemical, Ann Arbor, MI; 1:700), rabbit anti-MMP-3 (AB810; Chemicon International, Temecula, CA; 1:1000), mouse anti-GFAP (G3893; Sigma-Aldrich, Saint Louis, MO; 1:400), mouse anti-NeuN (MAB-377; Chemicon International, Temecula, CA; 1:400), and mouse anti-OX-42 (MAS370p; Accurate Chemical Corp., Westbury, NY; 1:100). Secondary antibodies and dilutions were: goat anti-rabbit linked to orange-fluorescent Alexa Fluor® 546 dye (Cy3 filter) and goat anti-mouse labeled with Alexa Fluor® 488 dye (FITC filter). Both secondary antibodies were obtained from Invitrogen (Carlsbad, CA) and diluted 1:500.

Double Immunofluorescence was performed as previously described (Yang et al., 2007). Briefly, slides were dried for 30 min at 50°C followed by incubation with cold acetone for 5 min. This was followed by two washes in PBS containing 0.1% Tween-20 (PBST). Nonspecific binding sites were blocked for 1 h at room temperature with PBST containing 1% bovine serum albumin (BSA; Sigma-Aldrich, Saint Louis, MO) and 5% normal goat serum (Invitrogen, Carlsbad, CA). Slides were incubated overnight at 4°C with primary antibodies. The next day, slides were washed three times in PBST and then incubated for 90 min at room temperature with secondary antibodies. Slides were placed in a 50 mM $NH_4Cl$ solution for 10 min to reduce non-specific fluorescence background. In some experiments, slides were incubated with 4'-6-diamidino-2-phenylinidole (DAPI) (Invitrogen, Carlsbad, CA) prior to coverslipping to label cell nuclei. Negative controls were incubated without the primary antibodies or with normal (non-immune) IgGs from the host species of the primary antibodies and failed to exhibit specific immunolabeling (not shown). All slides were viewed on an Olympus BX-51 fluorescence microscope (Olympus America Inc., Center Valley, PA) equipped with an Optronics digital camera and Picture Frame image capture software (Optronics, Goleta, CA). Dual immunofluorescence slides were also imaged confocally to verify co-labeling in a laser scanning confocal microscope (Zeiss LSM 510, Carl Zeiss Microimaging Inc., Thornwood, NY).

**Magnetic Resonance Imaging (MRI)**

For the MRI study, 8 rats were randomly divided into two groups: the first group consisting of 4 rats that received stereotaxic injection of rhTNF-$\alpha$ and treatment with indomethacin, while the second group of rats were injected with rhTNF-$\alpha$, but instead, were treated with the vehicle (n=4). After 24 h, rats were transported to the MRI room, placed in a dedicated rat holder and moved to the isocenter of the magnet prior to the imaging session. MR imaging was performed on a 4.7T Biospec® dedicated research MR scanner (Bruker Biospin, Billerica, MA), equipped with 500 mT/m (rise time 80-120 $\mu$s) gradient set (for performing small animal imaging) and a small bore linear RF coil (ID 72 mm) (Sood et al., 2007b). Initial localizer images were acquired using the following parameters: 2D FLASH (Fast Low Angle SHot), TR/TE 10/3 ms, matrix 256 x 128, FOV 6.4 cm, 1 slice per orientation. After the localizer images were acquired, T2 weighted and diffusion weighted imaging (DWI) was performed with the following parameters; T2 weighted-2D RARE (Rapid Acquisition with Relaxation Enhancement), TR/TE 4000/65 ms, FOV 3.2 cm x 3.2 cm, slice thickness 2 mm, slice gap 1 mm, number of slices 2, matrix 256 x 128, number of averages 5, receiver bandwidth 100 kHz; DWI - 2D Diffusion weighted RARE, TR/TE 2000/31.2 ms, FOV 3.2 cm x 3.2 cm, slice thickness 2 mm, slice gap 1 mm, number of slices 2, matrix 64 x 64, number of averages 5, receiver bandwidth 100 kHz, d = 5 ms, D= 20 ms, b value of 0 and 900 s $mm^{-2}$, diffusion gradient along the slice select direction.

The slices containing the lesion were identified from the T2 weighted structural images and same slice location was prescribed for all the subsequent MR protocols. MR data for generating T2 maps was acquired using the following parameters: 2D MSME (Multi Slice Multi Echo), TR 3000, TE1 30 ms, TE2 60 ms, TE3 90 ms, TE4 120 ms, FOV 3.2 cm x 3.2 cm, slice thickness 2 mm, slice gap 1 mm, number of slices 2, matrix 64 x 64, number of averages 4, receiver bandwidth 100 kHz.

The MR protocol for acquiring data for the Patlak plot method was then implemented (Patlak et al., 1983). In this acquisition, a reference baseline acquisition using the fast T1 mapping protocol was obtained. Rats were injected 0.2 mmol/kg of gadolinium-diethylenetriaminepentaacetic acid (Gd-DTPA, MW= 938 Da; Berlex, Montville, NJ) as a bolus into the femoral vein via an indwelling catheter, followed by imaging with rapid T1 mapping protocol, collecting 14 images over 45 minutes. The following optimized MR imaging parameters were used for this protocol: axial plane, 2D IR-SE-EPI, TR/TE 8s/19.4 ms, FOV 4.0 cm x 4.0 cm, slice thickness 2 mm, slice gap 1 mm, number of slices 2, matrix 64 x 64, number of averages 2, receiver bandwidth 250 kHz. Non slice selective magnetization inversion was performed using a hyperbolic secant (sech) RF pulse with pulse width 4 ms. The T1 mapping specific parameters for this protocol were; time for inversion (TI) TI = (100 + 600 x n) ms where n = 0, 1, 2, …12, number of TI points (n) is 13. The total scan time was 3 minutes and 12 seconds for each time point. The MR protocol parameters for rapid T1 mapping were optimized for accuracy of T1 relaxation time estimate in a single, normal rat brain (control) prior to including that animal in the study. The acquired data were transferred to a dedicated computer workstation for post processing. Post processing of the raw data involved generating Apparent Diffusion Coefficient (ADC) maps from DWI images, T2 maps, T1 maps from the raw data, reconstruction of permeability coefficient maps and construction of the Patlak plots.

All the data processing was performed using in-house software written in 64-bit MATLAB (Mathworks, Natick, MA) and implemented on a 64-bit processor (AMD64) workstation running Red Hat Enterprise Linux v3 (64-bit). Image analysis was performed using ImageJ (NIH, Bethesda, MD) and MRVision (Winchester, MA) software. T2 maps are gray scale images with pixel signal intensity values proportional to T2 relaxation times and shades of gray scaled to corresponding pixel-wise T2 values. A region of Interest (ROI) was drawn on T2 and ADC maps around the lesion on the ipsilateral side (TNF-$\alpha$ injected) and a similar ROI was drawn on the contralateral (normal) side. In the permeability coefficient maps reconstructed using MRI data, pixel signal intensity values are proportional to permeability coefficient values. Pixels with high and low signal intensity correspond to pixels with high and low permeability coefficient values respectively. Before analyzing the permeability maps i.e. drawing ROI, intracranial volume was differentiated from skull and image background by thresholding signal intensity from recognized anatomical structures identified on DW and T2w images. Image analysis was performed on permeability coefficient maps using the software described above. The permeability map for each slice was matched with DW image and ADC map to identify normal and injured regions. The lesion was identified on T2w and DW images as regions with long T2 relaxation time and low ADC values respectively. ROI was drawn manually by a single operator around regions with high signal intensity on the ipsilateral side and the same ROI was placed on the contralateral side. This step was repeated twice by the same operator in order to reduce intra-user bias. Analysis was repeated for the data sets obtained from vehicle and indomethacin-treated rats.

**Fluorometric Assay of Matrix Metalloproteinase-3 (MMP-3) Enzymatic Activity**

The activity of MMP-3 (stromelysin-1) was measured fluorometrically using a 5-FAM/QXL™520 fluorescence resonance energy transfer (FRET) peptide (Cat. No. 60580; AnaSpec Inc., San Jose, CA). In the intact FRET peptide (5-FAM-Arg-Pro-Lys-Pro-Val-Glu-Nva-Trp-Arg-Lys(QXL™520)-NH$_2$), the fluorescence of 5-FAM (5-carboxyfluorescein) is quenched by QXL™520. Upon cleavage into two separate fragments by the MMP-3 present in the sample, the fluorescence of 5-FAM is recovered, and can be monitored at excitation/emission wavelengths of 490/520 nm. This peptide has been documented to be cleaved by only MMP-3 and MMP-12 (macrophage elastase), but not by other MMPs (Nagase et al., 1994). Rat brain homogenates containing 100 µg of protein (aliquot from the lysate prepared for zymography) were mixed with assay buffer (50 mM Tris-HCl pH 7.6, 200 mM NaCl, 5 mM CaCl$_2$, 20 µM ZnSO$_4$ and 0.05% Brij-35) containing the FRET peptide (1 µM final concentration) to a final volume of 200 µl. The change in fluorescence (expressed as relative fluorescence units, RFU) was monitored at 5-min intervals for 1 h at room temperature using a luminescence spectrometer (PerkinElmer Instruments, Model LS55; Buckinghamshire, UK) attached to a workstation running FL WinLab software.

**Measurement of Reduced Glutathione (GSH)**

Levels of GSH were determined in rat brain homogenates using the monochlorobimane (MCB) method (Kamencic et al., 2000). MCB binds to GSH through interaction with glutathione-S-transferase (GST) to form a stable fluorescent adduct (MCB-GSH), allowing determination of GSH levels through fluorescent intensity of the MCB-GSH adduct. MCB has more selectivity for GSH over other thiols when compare with other bimanes, and was confirmed to react only with GSH (Fernandez-Checa and Kaplowitz, 1990). Brain homogenates (an aliquot from the homogenate prepared for zymography) containing 150 µg of protein were mixed, in duplicates, with the assay buffer (100 mM potassium phosphate buffer pH 6.5) containing 1 U/ml of GST from equine liver (BioVision, Mountain View, CA) and 0.5 mM of MCB (Sigma-Aldrich, Saint Louis, MO) in a black 96-well plate. Samples were incubated in the dark for 30 min at 37°C. The fluorescent product formed was quantified using a spectrofluorometer (PerkinElmer Instruments, Model LS55; Buckinghamshire, UK) at excitation and emission wavelengths of 380 and 460 nm, respectively.

Fluorescence in each well was recorded for 3 sec, with a slit width of 5 nm and a cut-off set at 430 nm. Tissue GSH concentration was determined using a GSH standard curve and expressed as µg GSH/mg protein.

**Drug Treatment**

The COX inhibitors used in the present study included nimesulide (COX-2 selective), Valeroyl Salicylate (VAS; COX-1 selective) and indomethacin (non-selective). Drugs were obtained from Cayman Chemical (Ann Arbor, MI), and dissolved in 5% Cremophor® EL (Sigma-Aldrich, Saint Louis, MO) in saline for intraperitoneal (i.p.) administration. The doses used were 12, 20 and 10 mg/kg for nimesulide, VAS and indomethacin, respectively. Animals were treated with the COX inhibitors immediately after rhTNF-$\alpha$ injection and again at 8 h. These doses and treatment paradigms have been previously demonstrated to be effective in reducing PGE$_2$ levels in the rodent brain (Paoletti et al., 1998; Candelario-Jalil et al., 2003c; Candelario-Jalil et al., 2007a).

**Statistical Analysis**

Data were analyzed using an unpaired *t*-test (2 groups) or one-way analysis of variance (ANOVA) with a *post-hoc* Student-Newman-Keuls test (multiple comparisons). In all statistical tests, differences were considered significant when p<0.05. Data are presented as mean ± standard error of the mean (S.E.M). Statistical analysis was performed using the program GraphPad Prism for Windows, Version 4.0 (GraphPad Software Incorporated, San Diego, CA).

**RESULTS**

**Inhibition of COX with indomethacin or VAS protects against rhTNF-$\alpha$-induced BBB opening in the rat**

Using the $^{14}$C-sucrose method we found that the intracerebral injection of rhTNF-$\alpha$ produced a significant increase in the BBB opening at 24 h (**Fig. 1**). We then investigated the effects of the administration of different COX inhibitors on BBB permeability changes induced by rhTNF-$\alpha$. Using selective inhibitors of each of the COX isozymes, we found that indomethacin and the COX-1 selective inhibitor VAS significantly (p<0.05) reduced the BBB opening in this animal model of neuroinflammation. However, the COX-2 selective inhibitor nimesulide failed to significantly reduce rhTNF-$\alpha$-induced BBB breakdown (**Fig. 1**).

**Figure 2A** shows a series of T2 weighted images and color-coded permeability coefficient maps for two representative rats: rhTNF-$\alpha$ plus vehicle (top row) and rhTNF-$\alpha$ plus indomethacin (bottom row). The T2 weighted images demonstrated the lesion in relation to the anatomical detail in the rat brain. On the T2 maps, a 12% reduction in T2 values was observed on the ipsilateral lesion side in the indomethacin treated animals compared to the rhTNF-$\alpha$ plus vehicle group. There was no significant reduction in Apparent Diffusion Coefficient (ADC) values in the indomethacin-treated group compared to the rhTNF-$\alpha$ plus vehicle group. Permeability coefficient color maps showed a reduction below $1.0 \times 10^{-3}$ ml g$^{-1}$ min$^{-1}$ in the indomethacin treated group compared to over $3.0 \times 10^{-3}$ ml g$^{-1}$ min$^{-1}$ in the rhTNF-$\alpha$ plus vehicle group (note the scale differences). Figure 2B shows a graph of mean permeability coefficient values in rhTNF-$\alpha$ plus indomethacin group and rhTNF-$\alpha$ plus vehicle estimated using MRI. In the rhTNF-$\alpha$ plus vehicle group, the estimates in the ipsilateral and contralateral hemisphere were $5.47 \pm 1.0 \times 10^{-3}$ ml g$^{-1}$ min$^{-1}$ and $0.38 \pm 0.26 \times 10^{-3}$ ml g$^{-1}$ min$^{-1}$, respectively. The permeability coefficient estimates in rhTNF-$\alpha$ plus indomethacin group were $0.41 \pm 0.18 \times 10^{-3}$ ml g$^{-1}$ min$^{-1}$ and $0.31 \pm 0.18 \times 10^{-3}$ ml g$^{-1}$ min$^{-1}$ in the ipsilateral and contralateral hemisphere, respectively. On applying one-way ANOVA test, there was a significant difference in permeability coefficient values between the groups. Furthermore, a significant difference ($p<0.05$) in the permeability coefficient values between the lesion and non lesion side in rhTNF-$\alpha$ plus vehicle rats was observed. However, the mean permeability coefficient estimates between the ipsilateral and contralateral side in rhTNF-$\alpha$ plus indomethacin rats were not significantly different. There was a significant ($p<0.001$) reduction observed in mean permeability coefficient values on the ipsilateral side in the indomethacin treated rats as compared to the untreated rats.

**The dramatic increase in MMP-9 activity following rhTNF-$\alpha$ injection is reduced by indomethacin treatment**

Intracerebral injection of 325 ng (5000 U) of rhTNF-$\alpha$ produced a significant increase in MMP-9 activity after 24 h compared with the saline-injected control, as revealed by gelatin zymography (**Fig. 3A**). Inhibition of COX with indomethacin produced a significant reduction in the gelatinolytic activity of MMP-9 (**Fig. 3B**). No significant change in MMP-2 activity was observed in any of the treatment groups as compared to saline-injected animals.

**Indomethacin limits the induction of MMP-9 protein expression following intracerebral injection of rhTNF-$\alpha$**

In our next experiments, we used immunoblotting to investigate the effect of indomethacin on the expression of MMP-9, MMP-2, COX-1 and COX-2 at 24 h following the injection of rhTNF-$\alpha$ into the rat striatum. MMP-9 protein levels increased dramatically in rhTNF-$\alpha$-injected rats (**Fig. 4A**). There was a significant reduction (p<0.05) in MMP-9 protein expression in animals injected with rhTNF-$\alpha$ and given indomethacin, as compared with vehicle-treated controls (**Fig. 4A** and **4B**). In addition, the expression of COX-2 was significantly increased in the brain of rats injected with rhTNF-$\alpha$ when compared with saline injection (**Fig. 4A** and **4C**). However, indomethacin treatment had no significant effect on rhTNF-$\alpha$-mediated COX-2 induction (**Fig. 4C**). Similar to the findings of the gelatin zymography analysis, rhTNF-$\alpha$ did not produce any increase in MMP-2 protein expression (**Fig. 4D**), and indomethacin administration did not significantly affect MMP-2 protein levels (**Fig. 4A** and **4D**). Furthermore, COX-1 expression was unaffected by rhTNF-$\alpha$ and did not change in rats administered indomethacin (**Fig. 4 A** and **4E**).

**COX inhibition with indomethacin reduces rhTNF-$\alpha$-induced MMP-3 protein expression and activity**
Since stromelysin-1 (MMP-3) contributes to BBB disruption during lipopolysaccharide (LPS)-induced neuroinflammation (Gurney et al., 2006b), we investigated the effects of COX inhibition with indomethacin on MMP-3 expression. Rat pro-MMP-3 was detected by Western blot in its glycosylated form at approximately 62 kDa. Results from this experiment indicated that pro-MMP-3 expression is dramatically induced by rhTNF-$\alpha$ when compared to saline (**Fig. 5A** and **5B**). Very low levels of pro-MMP-3 were observed in saline-injected animals, and this was only observed when exposing the membrane to autoradiography film for longer periods of time (not shown). Treatment with indomethacin markedly reduced (p<0.05) pro-MMP-3 protein levels in comparison with TNF-$\alpha$ plus vehicle (**Fig. 5A** and **5B**). Fluorescent enzymatic assay for MMP-3 activity showed a significant increase (p<0.01) in MMP-3 activity in the rhTNF-$\alpha$ plus vehicle group in relation to saline control, which was significantly (p<0.05) reduced by treatment with indomethacin (**Fig. 5C**).

**Cellular sources of MMP-9, MMP-3 and COX-2 following intracerebral injection of rhTNF-$\alpha$**
Since intracerebral injection of rhTNF-$\alpha$ produced a significant increase in MMP-9, MMP-3 and COX-2 levels, as revealed by the Western blot experiments (**Figs. 4 and 5**), we were interested in identifying the cellular source of these inducible proteins following rhTNF-$\alpha$-mediated neuroinflammation. Double immunofluorescence labeling performed at 24 h showed that both, neurons (NeuN positive) and microglia/macrophages (OX-42 positive), were the cell types responsible for the enhanced MMP-9 production after rhTNF-$\alpha$ injection (**Fig. 6A**). Very few GFAP-positive astrocytes were immunoreactive for MMP-9 in this model. Further inspection of these brain sections under a laser scanning confocal microscope confirmed that MMP-9-expressing cells were NeuN and OX-42 positive (**Fig. 6A**). Double immunolabeling for MMP-3 and each specific cellular marker was conducted in saline- and rhTNF-$\alpha$- injected brains (treated with indomethacin or the vehicle). We observed colocalization of MMP-3 with NeuN- and OX-42-positive cells, but failed to find MMP-3 immunoreactivity in GFAP-positive astrocytes (**Fig. 6B**). A similar immunolabeling analysis was performed for COX-2. We observed COX-2 immunoreactivity in neurons and microglia/macrophages (not shown). Compared to rhTNF-α-injected brains, saline-injected controls showed very few cells positive for either MMP-9, MMP-3 or COX-2 (not shown).

**Indomethacin prevented GSH depletion during rhTNF-α-mediated neuroinflammation**

COX activity is an important source of oxidative stress during neuroinflammation (Candelario-Jalil et al., 2003b; Pepicelli et al., 2005; Akundi et al., 2005a). In order to investigate the effects of indomethacin treatment on endogenous antioxidant capacity following rhTNF-α, we measured GSH levels in brain homogenates prepared from saline, rhTNF-α plus vehicle, and rhTNF-α plus indomethacin groups. A significant reduction in GSH was found in the rhTNF-α plus vehicle group in relation to saline-injected animals and indomethacin reversed rhTNF-α-induced GSH depletion (**Fig. 7**).

**DISCUSSION**

Our results link MMPs to COX-mediated disruption of the BBB during TNF-α-induced neuroinflammation. We have demonstrated for the first time that pharmacological inhibition of COX with indomethacin protects against BBB disruption. Indomethacin significantly attenuated the activity and expression of MMP-9 and MMP-3, and prevented the loss of GSH following intracerebral injection of TNF-α.

Other investigators have shown, in an *in vitro* model of the BBB that enhanced COX activity, with the concomitant increase in prostaglandin formation, is involved in TNF-α-mediated increase in permeability. Inhibition of COX activity with indomethacin or with the COX-2 selective inhibitor NS-398 prevented TNF-α-induced changes in permeability (Mark et al., 2001a).

A better understanding of the molecular mechanisms involved in TNF-α-mediated increase in MMPs could lead to the discovery of new potential targets amenable to pharmacological manipulation in order to limit BBB opening and brain edema in many neurological disorders. Earlier we showed that MMP-9 was induced by intracerebral injection of TNF-α (Rosenberg et al., 1995a). MMP blockade with the MMP inhibitor, BB-94 (Batimastat), prevented the BBB breakdown seen at 24 h following TNF-α injection (Rosenberg et al., 1995c), suggesting that enhanced MMP-9 activity is pivotal in BBB disruption. Present data extend those findings by demonstrating that increased BBB permeability induced by TNF-$\alpha$ appears to be linked to the increase in MMP-9 and MMP-3 through a COX-dependent mechanism. Furthermore, we identified for the first time the cellular source of MMPs in this model of neuroinflammation. Double immunofluorescence and confocal microscopy showed that neurons and microglia/macrophages produce large amounts of MMP-9, COX-2 and MMP-3 following injection of TNF-$\alpha$. The colocalization of MMP-9 and MMP-3 in OX-42-positive cells around blood vessels is indicative of their role in BBB damage induced by TNF-$\alpha$.

The increased MMP-3 expression and activity observed after TNF-$\alpha$ represents the first *in vivo* evidence of the effect of this pro-inflammatory cytokine on MMP-3 levels in the injured brain. In addition, our data show that indomethacin treatment significantly reduces MMP-3 levels and enzymatic activity following TNF-$\alpha$-mediated neuroinflammatory events. These results are in line with an earlier *in vitro* study showing the protective efficacy of indomethacin against a TNF-$\alpha$-induced increase in MMP-3 in fibroblasts (Morin et al., 1999).

Previous investigations have demonstrated that both MMP-3 and free radicals are involved in MMP-9 activation during inflammatory events (Ogata et al., 1992; Gasche et al., 2001; Rosenberg et al., 2001; Gu et al., 2002). In addition, MMP-3 is able to cleave most components of the extracellular matrix (Wilhelm et al., 1987). Recently, we showed that MMP-3 is involved in BBB injury, neutrophil infiltration, and degradation of tight junction proteins following administration of LPS into the brain (Gurney et al., 2006a).

Using gelatin zymography to measure MMP activity in TNF-$\alpha$-injected brains only demonstrated latent forms of MMP-2 and MMP-9. Enzymatic activation of the gelatinases occurs at the cell surface and the activated species is quickly degraded resulting in low levels of active MMP in the tissue (Fridman et al., 2003b). In addition, activation of MMP-9 can occur without removal of the inhibitory pro-peptide domain and consequently generate enzymatic activity in the absence of a noticeable change in molecular mass (Fridman et al., 2003a).

Although the COX/PGE$_2$ pathway is important in cancer and peripheral inflammation by enhancing MMP production, little is known about this pathway in neuroinflammation and brain injury. In an earlier study, a COX-2 selective inhibitor, NS-398, significantly reduced MMP-9 mRNA expression following radiation brain injury (Kyrkanides et al., 2002). Our finding that pharmacological inhibition of COX significantly reduced MMP-3 and MMP-9 protein expression and enzymatic activity in an *in vivo* model of brain injury could explain the efficacy of COX inhibitors in protecting against BBB disruption following cerebral ischemia (Ting, 1990; Zuckerman et al., 1994; Candelario-Jalil et al., 2007f), bacterial meningitis (Kadurugamuwa et al., 1989), and against edema associated with brain tumors (Badie et al., 2003).

The availability of selective inhibitors of the COX isozymes provides a powerful pharmacological tool to dissect the relative contribution of each isoform to the inflammatory process. By using this approach in the present study, we found that COX-1 seems to be more important than COX-2 in mediating TNF-$\alpha$-induced BBB permeability changes. This finding is in contrast with results from our recent study in a rat model of focal cerebral ischemia, where COX-2 inhibition with nimesulide, but not COX-1 inhibition with VAS, potently reduced BBB breakdown in the ischemic cerebral cortex (Candelario-Jalil et al., 2007c). The apparent discrepancies among our studies, in terms of which COX isoform is playing the most important role in BBB damage, could be possibly explained by the different animal models employed (ischemia *versus* TNF-$\alpha$). In addition, in the focal ischemic model we found that COX-2 inhibition failed to reduce BBB disruption in the striatum, and no significant increase in PGE$_2$ levels were observed in the striatal homogenates prepared from ischemic animals as compared to sham controls (Candelario-Jalil et al., 2007d). In our present study, TNF-$\alpha$ was injected intrastriatally, therefore we might speculate that the protective effect of COX-2 inhibition against BBB opening is region-specific. In an earlier study, we found that *both* COX isoforms contribute to oxidative damage and neurodegeneration in the hippocampus following a global cerebral ischemic event (Candelario-Jalil et al., 2003a). Thus, it seems that the contribution of each COX isozyme to the neuroinflammatory process depends on the type of injury, brain region under study, and possibly other unknown factors.

It is interesting that, despite being upregulated by TNF-$\alpha$, COX-2 does not seem to mediate BBB disruption induced by this pro-inflammatory cytokine. It is possible that it takes several hours for a catalytically active COX-2 protein to be present in the injected site. By then, most of the free arachidonic acid released by the TNF-$\alpha$-induced PLA$_2$ activity has been metabolized by COX-1. In this particular scenario following acute injection of TNF-$\alpha$, arachidonic acid metabolism through COX-1 seems to play a greater role than through COX-2. This notion is supported by our findings that COX-1, but not COX-2, inhibition significantly limits TNF-$\alpha$-mediated BBB breakdown. However, the observation that indomethacin confers more protection than the COX-1 inhibitor VAS may suggest that the relative small contribution of COX-2 should not be ignored.

The differences we observed between ischemia- and TNF-$\alpha$-induced BBB openings, in terms of which COX isoform is involved, may have important implications for the potential therapeutic use of COX

*15*

inhibitors to limit the BBB opening in different clinical conditions. Non-selective COX inhibitors such as indomethacin might be useful in the management of BBB breakdown following inflammatory processes (e.g., bacterial meningitis), whereas in cerebral ischemia, a selective COX-2 inhibitor would be preferable (Candelario-Jalil et al., 2007e). Thus, the present findings herein shed more light into the specific role of COX in brain injury, and might have relevance for the potential use of COX inhibitors in the treatment of neurological disorders associated with BBB disruption.

T2 maps provide a useful way of quantifying edema using MRI. A 12% reduction in T2 values from T2 maps in the indomethacin treated group suggests that there was a decrease in edema due to drug treatment. This could probably be explained by the anti-inflammatory action of indomethacin. The reduction in permeability coefficient values due to indomethacin treatment was found to be statistically significant ($p<0.05$) and confirms the findings from the $^{14}$C-sucrose technique. This suggests that indomethacin may be working to reduce BBB leakage in the ipsilateral region and reiterates the BBB blocking effect of indomethacin and also suggests that indomethacin affects permeability on the ipsilateral side only and has no affect on the permeability in healthy tissue.

In summary, the present study provides compelling evidence implicating COX activity in the TNF-$\alpha$-induced BBB disruption. Inhibition of COX with indomethacin prevented BBB opening through a mechanism which possibly involves reduction in MMP-3 and MMP-9 expression/activity and prevention of the loss of endogenous antioxidant capacity. This investigation adds significant information on the molecular mechanisms underlying the protection seen with COX inhibitors against BBB damage in several models of brain injury. Our results show for the first time that BBB disruption during neuroinflammation can be significantly reduced by administration of COX inhibitors. We propose that modulation of COX in brain injury by COX inhibitors or agents modulating prostaglandin $E_2$ formation/signaling may be useful in the clinical setting where the BBB is compromised by various pathologies.

**ACKNOWLEDGMENTS**

We thank Dr. Lee Anna Cunningham (Department of Neuroscience, University of New Mexico, USA) for her critical comments on the manuscript. Authors are grateful to Dr. Qing-Xiang Amy Sang (Department of Chemistry and Biochemistry, Florida State University, Tallahassee, FL, USA) for advice and help with the MMP-3 activity assay.

# References

Akundi RS, Candelario-Jalil E, Hess S, Hull M, Lieb K, Gebicke-Haerter PJ and Fiebich BL (2005b) Signal transduction pathways regulating cyclooxygenase-2 in lipopolysaccharide-activated primary rat microglia. *Glia* **51:**199-208.

Akundi RS, Candelario-Jalil E, Hess S, Hull M, Lieb K, Gebicke-Haerter PJ and Fiebich BL (2005a) Signal transduction pathways regulating cyclooxygenase-2 in lipopolysaccharide-activated primary rat microglia. *Glia* **51:**199-208.

Badie B, Schartner JM, Hagar AR, Prabakaran S, Peebles TR, Bartley B, Lapsiwala S, Resnick DK and Vorpahl J (2003) Microglia cyclooxygenase-2 activity in experimental gliomas: possible role in cerebral edema formation. *Clin Cancer Res* **9:**872-877.

Candelario-Jalil E, Gonzalez-Falcon A, Garcia-Cabrera M, Alvarez D, Al-Dalain S, Martinez G, Leon OS and Springer JE (2003a) Assessment of the relative contribution of COX-1 and COX-2 isoforms to ischemia-induced oxidative damage and neurodegeneration following transient global cerebral ischemia. *J Neurochem* **86:**545-555.

Candelario-Jalil E, Gonzalez-Falcon A, Garcia-Cabrera M, Alvarez D, Al-Dalain S, Martinez G, Leon OS and Springer JE (2003b) Assessment of the relative contribution of COX-1 and COX-2 isoforms to ischemia-induced oxidative damage and neurodegeneration following transient global cerebral ischemia. *J Neurochem* **86:**545-555.

Candelario-Jalil E, Gonzalez-Falcon A, Garcia-Cabrera M, Alvarez D, Al-Dalain S, Martinez G, Leon OS and Springer JE (2003c) Assessment of the relative contribution of COX-1 and COX-2 isoforms to ischemia-induced oxidative damage and neurodegeneration following transient global cerebral ischemia. *J Neurochem* **86:**545-555.

**FOOTNOTES**

*Financial Support*

This study was partially supported by a grant from the National Institutes of Health (NIH) to GAR (5RO1NS04547). ECJ was supported by a grant from the American Heart Association (0720160Z, Pacific Mountain Affiliate). Confocal images in this study were generated in the University of New Mexico Cancer Center Fluorescence Microscopy Facility, which received support from NCRR 1S10 RR14668, NSF MCB9982161, NCRR P20 RR11830, NCI P30 CA118100, NCRR S10 RR19287, NCRR S10 RR016918, the University of New Mexico Health Sciences Center, and the University of New Mexico Cancer Center.

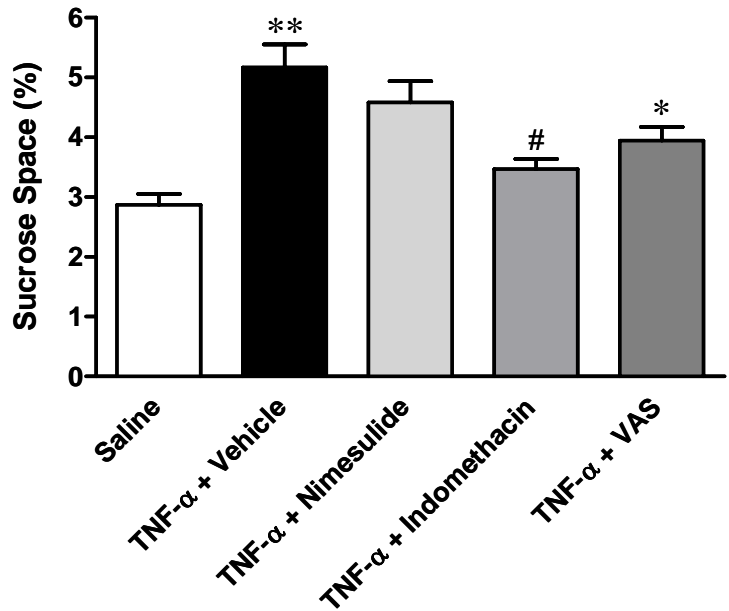

**Fig. 1.** Effects of COX inhibitors on the uptake of $^{14}$C-sucrose into rat brain as a measure of BBB permeability. Rats received an intrastriatal injection of 5000 U (325 ng) of rhTNF-$\alpha$ and were treated with either nimesulide (COX-2 inhibitor; 12 mg/kg, i.p., n=6), Valeroyl salicylate (VAS; COX-1 selective inhibitor; 20 mg/kg, i.p., n=6), indomethacin (non-selective COX inhibitor; 10 mg/kg, i.p., n=6) or the vehicle (5% Cremophor® EL in saline, n=12). A control group included rats injected intracerebrally with saline and treated with the vehicle (n=8). Drugs were given immediately after rhTNF-$\alpha$ injection and again at 8 h. After 24 h of the rhTNF-$\alpha$ injection, rats were given 10 $\mu$Ci of $^{14}$C-sucrose intravenously and sacrificed 10 min later. Samples from striatum and blood were collected, and brain sucrose uptake was calculated as a percentage of sucrose in brain to that in blood (Sucrose Space in %). Indomethacin and VAS significantly reduced rhTNF-$\alpha$-induced BBB disruption. **$p<0.001$ with respect to saline. #$p<0.01$ and *$p<0.05$ with respect TNF-$\alpha$ + vehicle.

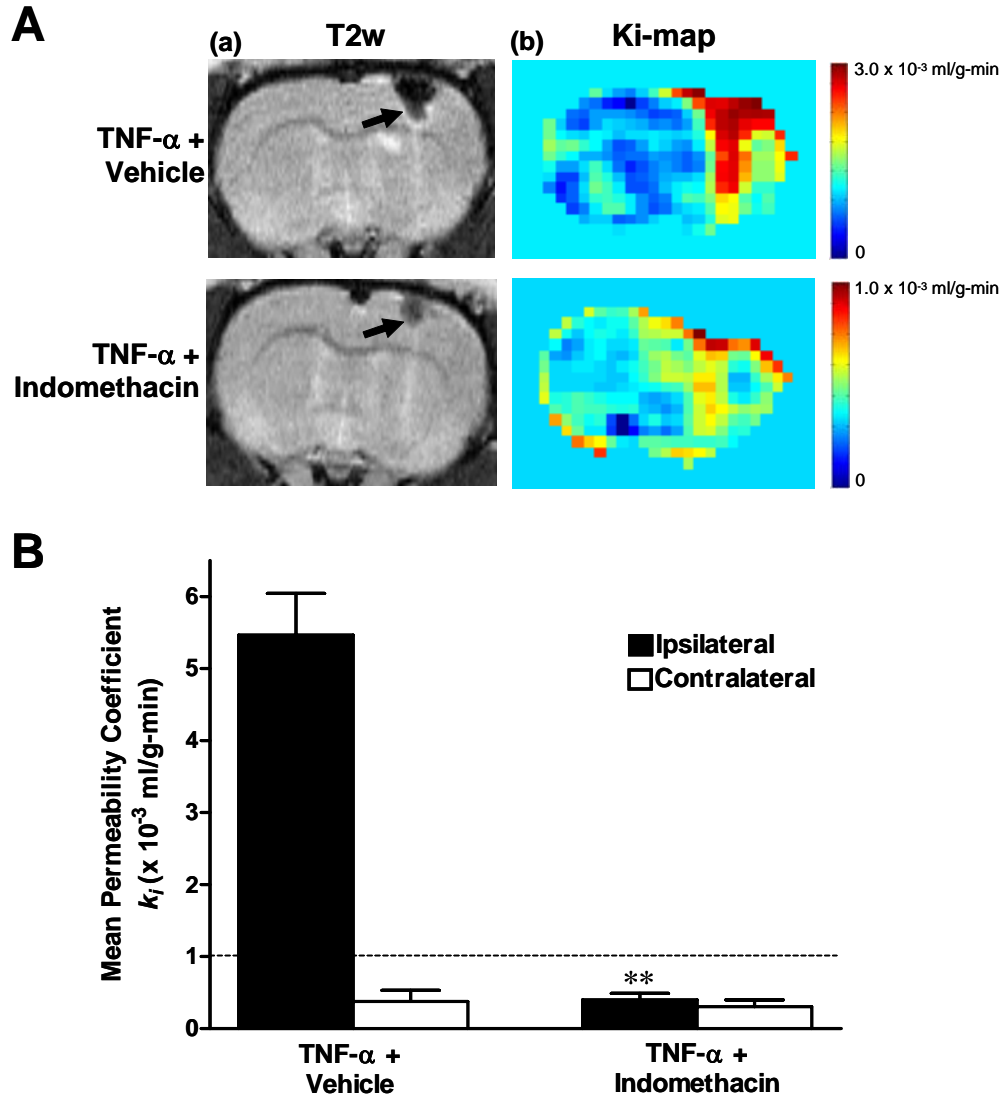

**Fig. 2. Panel A** shows **(a)** T2 weighted image and **(b)** color coded permeability coefficient map for TNF-α + vehicle and TNF-α + indomethacin group. The anatomical T2 weighted images clearly demonstrate the extent of the lesion in both untreated and treated groups. However, visually the extent of the lesions appears to be limited in the treated group compared to the untreated group. Color-coded permeability maps demonstrate clearly the regions of high (arrow) and low permeability in treated and control rats. In the treated group, indomethacin affects permeability by blocking BBB leakage. Note the different color scales used for the permeability maps. **Panel B:** A plot of mean permeability coefficient estimates in TNF-α + vehicle (n=4) and TNF-α + indomethacin group (n=4) rats obtained using the MRI technique. The dashed line represents the upper limit of the range (0-1 x $10^{-3}$ ml/g-min) of permeability coefficient values in healthy tissue. Rats treated with indomethacin demonstrated a significant reduction in permeability coefficient values on the side of rhTNF-α injection as compared to the untreated rats. **p<0.05 with respect to vehicle ipsilateral. No significant difference was observed on the contralateral side between the control and treated rats. Bars represent mean ± S.E.M.

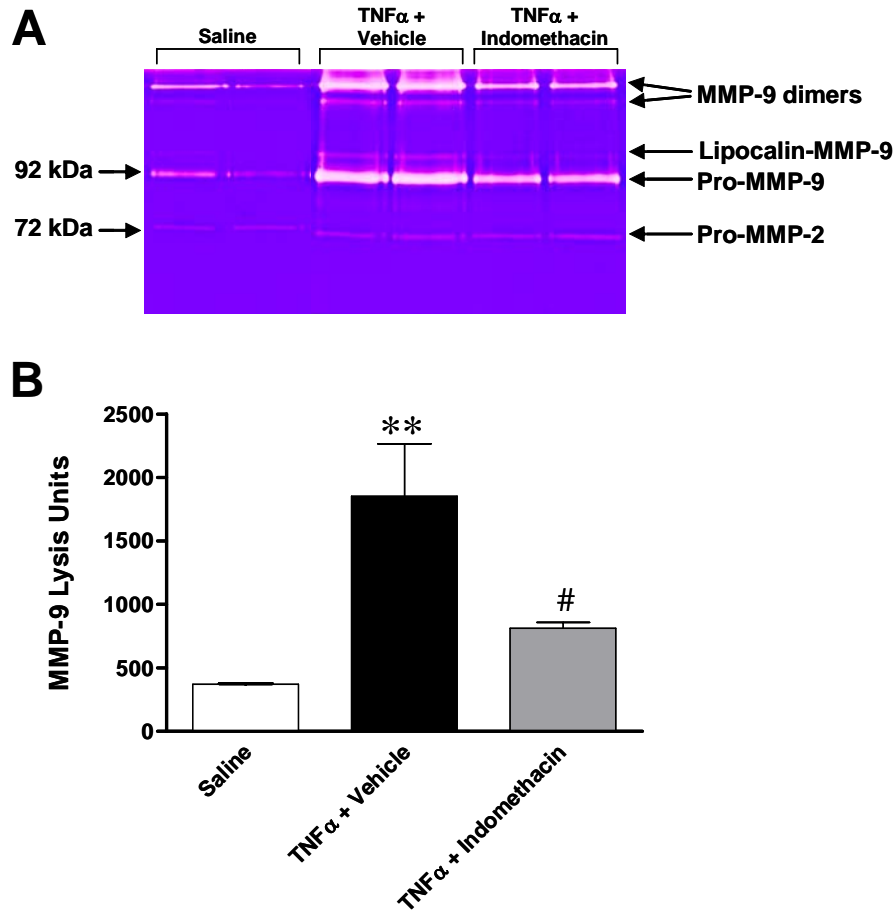

**Fig. 3. A:** Representative gelatin zymogram showing the significant reduction by indomethacin of rhTNF-α-mediated increase in MMP-9 activity. Animals were injected in the striatum with either saline or 5000 U of rhTNF-α. Rats were treated with indomethacin (10 mg/kg, i.p., immediately after rhTNF-α and again after 8 h; n=6). Another group of animals was treated with the vehicle agent (n=6). Protein samples were analyzed by gelatin zymography to detect the presence of gelatinases at 24 h after rhTNF-α injection. **B:** Quantitative estimation of gelatinolytic activity of MMP-9 by gelatin zymography analysis. Densitometric and statistical analyses show increases in the activity of MMP-9 following rhTNF-α injection and a significant reduction in animals administered indomethacin. No significant increase in MMP-2 activity was found after rhTNF-α microinjection. Densitometric analysis of lytic zones at 72 kDa, corresponding to pro-MMP-2 activity, showed no significant changes among treatments (not shown). Bars represent mean ± S.E.M. of the 92-kDa band in arbitrary densitometric units. **p<0.05 with respect to saline. #p<0.05 with respect to the rhTNF-α plus vehicle group.

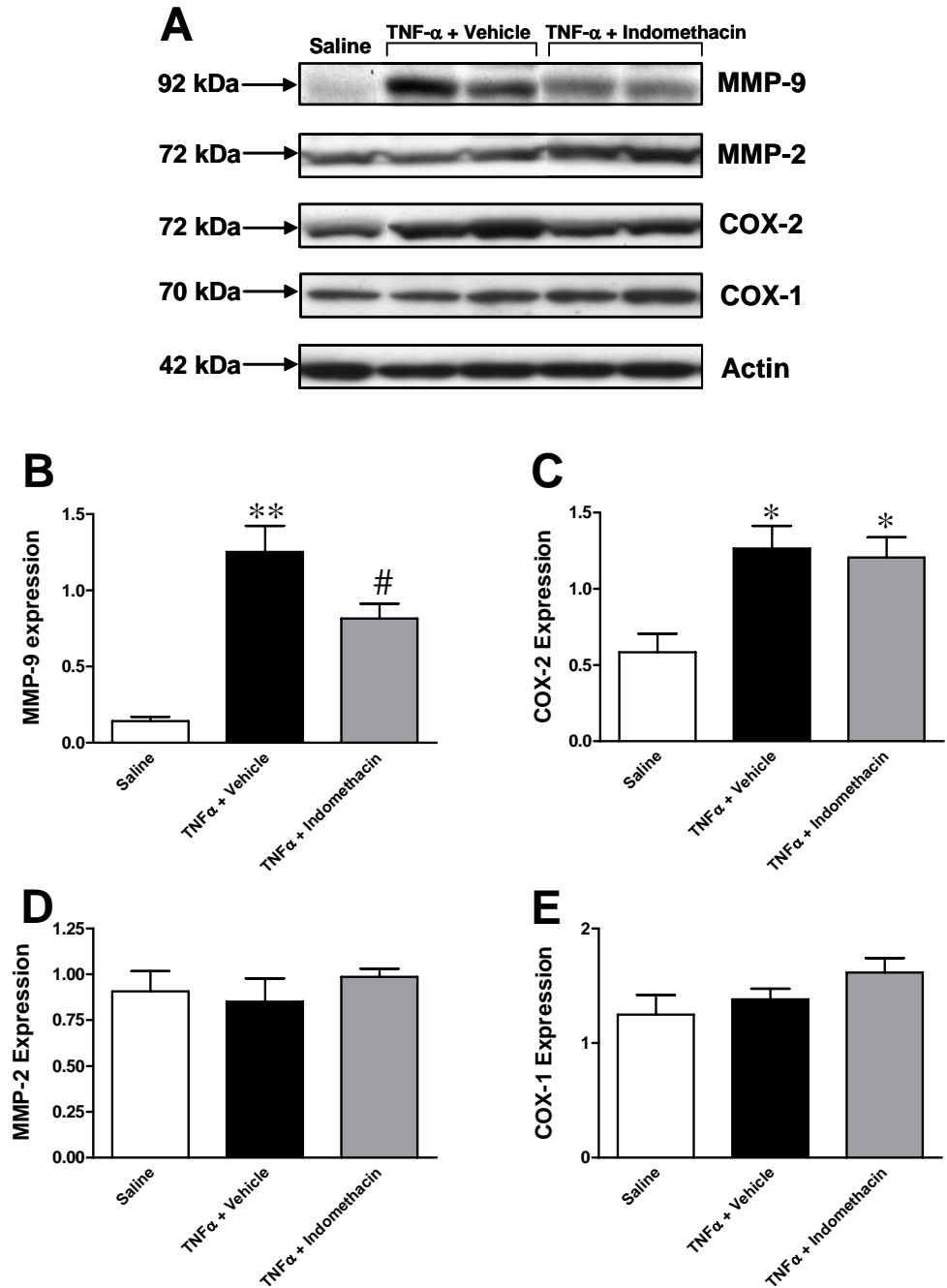

Fig. 4. Immunoblot analysis of protein levels of MMP-9, MMP-2, COX-2, COX-1 and actin in the rat brain after intracerebral injection of 325 ng of rhTNF-α (equivalent to 5000 U of TNF-α) or physiological saline. Animals were treated with the COX inhibitor indomethacin (10 mg/kg, i.p., immediately after rhTNF-α, and again after 8 h; n=6) or the vehicle agent (n=6). Protein extracts were prepared after 24 h of the injection with rhTNF-α, and subjected to SDS-PAGE followed by immunoblot analysis using specific antibodies. **A:** Representative Western blots for each protein in animals treated with saline, rhTNF-α plus vehicle, and rhTNF-α plus indomethacin. **B-E:** Quantitative densitometric analysis of relative protein expression normalized to actin loading control. Indomethacin significantly reduced MMP-9 expression induced by rhTNF-α. **$p<0.05$ with respect to MMP-9 expression in group injected with saline. #$p<0.05$ with respect to the rhTNF-α plus vehicle group. *$p<0.05$ with respect to COX-2 expression in saline-injected brains. Bars represent mean ± S.E.M.

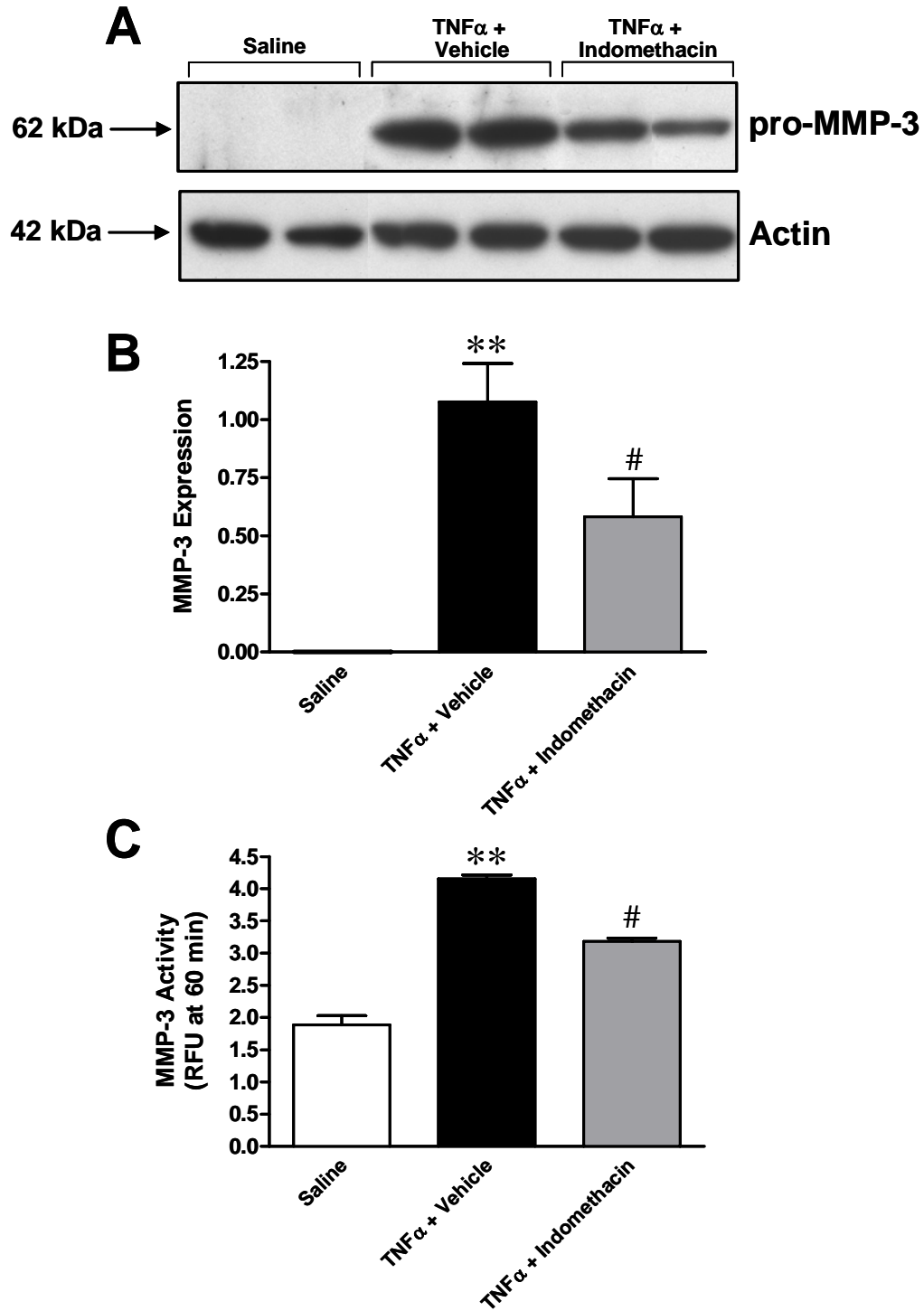

**Fig. 5. A:** Western blot analysis of protein levels of pro-MMP-3 and actin in the rat brain after intracerebral injection of 325 ng of rhTNF-$\alpha$ or physiological saline. Animals were treated with the vehicle agent (n=6) or with the COX inhibitor indomethacin (10 mg/kg, i.p., immediately after rhTNF-$\alpha$, and again after 8 h; n=6). Protein extracts were prepared after 24 h of the injection with rhTNF-$\alpha$, and subjected to SDS-PAGE followed by immunoblot analysis using specific antibodies. **Panel A** shows representative Western blots for each protein in animals treated with saline, rhTNF-$\alpha$ plus vehicle, and rhTNF-$\alpha$ plus indomethacin. **B:** Quantitative densitometric analysis of relative protein expression normalized to actin loading control. Indomethacin significantly reduced rhTNF-$\alpha$-induced pro-MMP-3 expression. Only very faint bands for active MMP-3 (~45 kDa) were seen by Western blot. These bands appeared after a relatively long exposure of the blot to the autoradiography film. This hampered the accurate quantification of these bands using densitometry due to enhanced background. Similarly, saline-injected brains showed faint pro-MMP-3 expression, which was seen only after extended exposure of the blot to X-ray film. **C:** Analysis of MMP-3 activity using a fluorescence resonance energy transfer (FRET) peptide. Significant increase of MMP-3 relative activity was seen in rhTNF-$\alpha$-injected brains as compared with saline injection. Treatment with the COX inhibitor indomethacin significantly reduced rhTNF-$\alpha$-induced MMP-3 activity in the 24-h brain lysates. **p$<$0.05 with respect to saline. #p$<$0.01 with respect to the rhTNF-$\alpha$ plus vehicle group. Bars represent mean ± S.E.M.

**A**

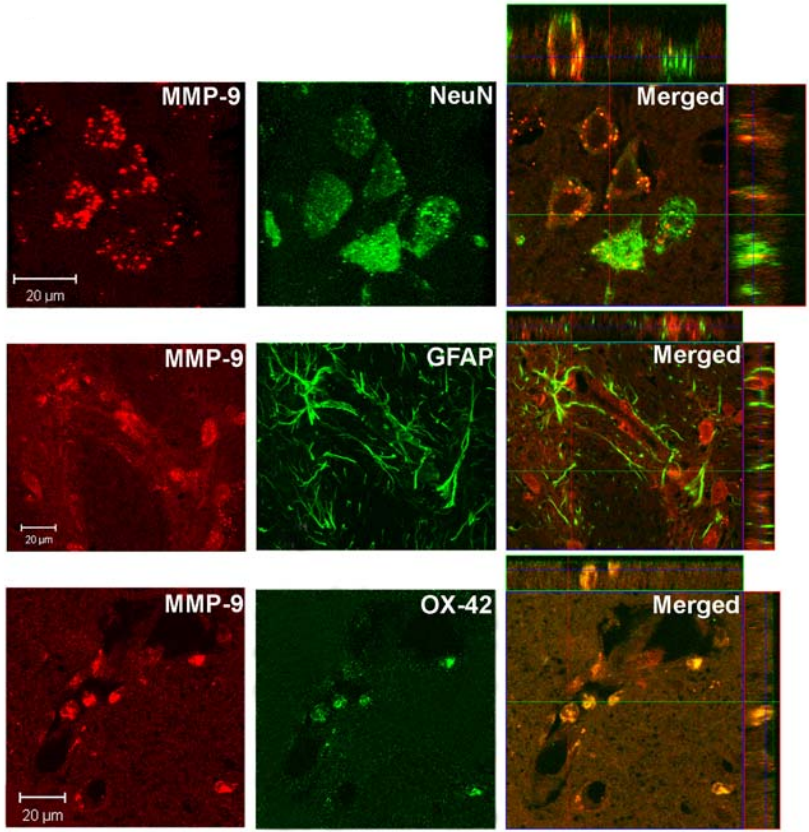

**B**

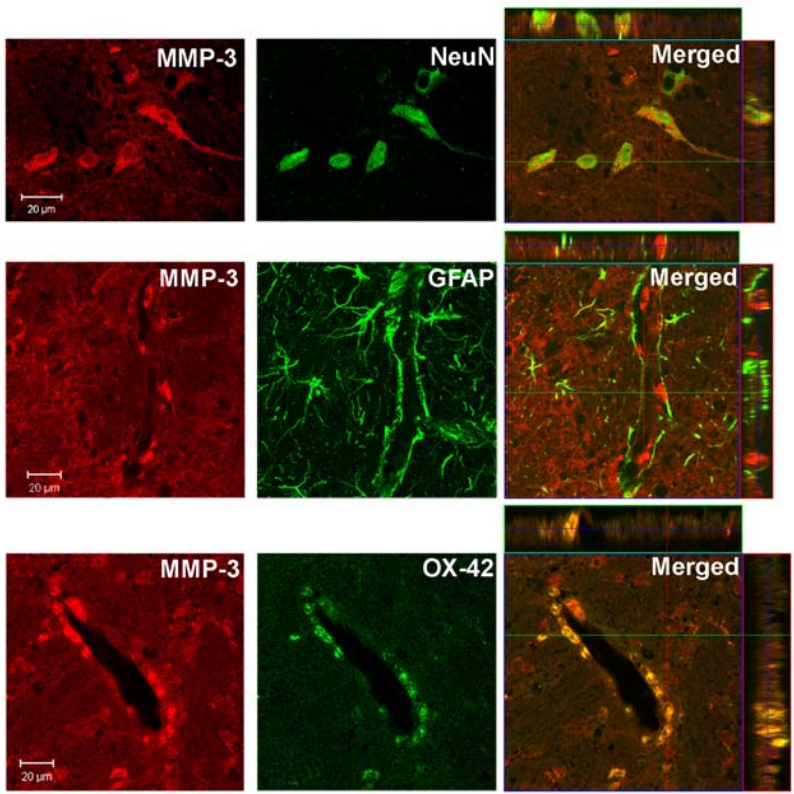

**Fig. 6.** Double immunofluorescence and confocal microscopy analysis showing the cellular source of MMP-9 and MMP-3 following intracerebral injection of rhTNF-α. **A:** Z-stack showing MMP-9 positive cells colocalize with the neuronal marker NeuN and with OX-42 positive cells (yellow color). Very few GFAP-positive astrocytes expressed MMP-9 upon injection of TNF-α. **B:** Confocal microscopy (Z-stack) confirmed the cellular source of MMP-3 to be neurons and OX-42 positive cells (microglia/macrophages). Most microglia/macrophages expressing MMP-3 and MMP-9 were seen around blood vessels.

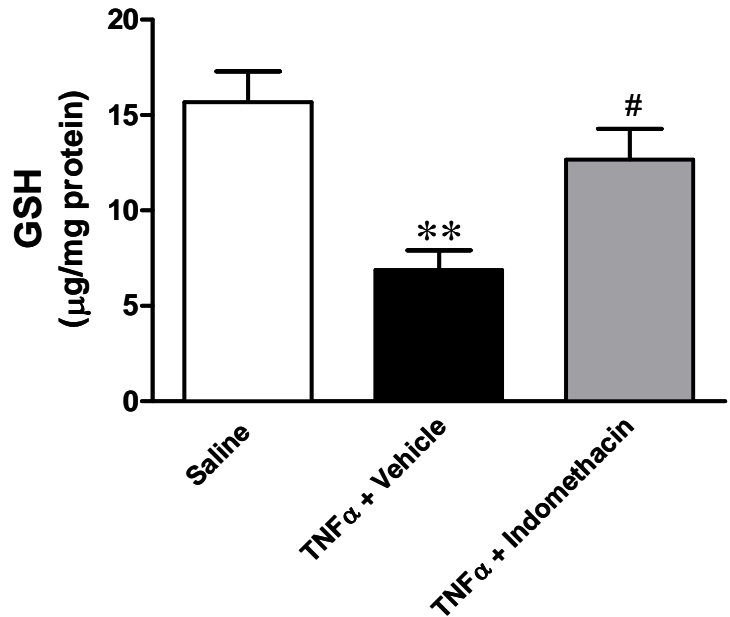

**Fig. 7.** Intracerebral injection of rhTNF-α produces a significant reduction in the levels of reduced glutathione (GSH) as compared to saline-injected control animals. Rats were treated with the vehicle agent (n=6) or with the COX inhibitor indomethacin (10 mg/kg, i.p., immediately after rhTNF-α, and again after 8 h; n=6). GSH levels were measured in brain homogenates after 24 h of the injection of rhTNF-α using the monochlorobimane (MCB) method. Homogenate extracts containing 150 μg of total protein were added to the assay buffer containing MCB and Glutathione-S-transferase and incubated for 30 min at 37°C in the dark. Formation of the MCB-GSH adduct was measured fluorometrically (Ex/Em=380/460 nm). Treatment with the COX inhibitor, indomethacin, prevented rhTNF-α-induced depletion of GSH. **$p<0.05$ with respect to saline. #$p<0.05$ with respect to the rhTNF-α plus vehicle group. Bars represent mean ± S.E.M.